\documentclass[12pt]{iopart}

\newtheorem{theorem}{Theorem}
\newtheorem{lemma}{Lemma}
\newtheorem{corollary}{Corollary}

\begin{document}

\title{Convergence theorems for quantum annealing}
\author{Satoshi Morita and Hidetoshi Nishimori}
\address{Department of Physics, Tokyo Institute of Technology, Oh-okayama,
Meguro-ku, Tokyo 152-8551, Japan}

\begin{abstract}
We prove several theorems to give sufficient conditions for convergence
of quantum annealing, which is a protocol to solve generic optimization
problems by quantum dynamics.  In particular the property of strong
ergodicity is proved for the path-integral Monte Carlo implementation of
quantum annealing for the transverse Ising model under a power decay of
the transverse field.  This result is to be compared with the much
slower inverse-log decay of temperature in the conventional simulated
annealing.  Similar results are proved for the Green's function Monte
Carlo approach.  Optimization problems in continuous space of particle
configurations are also discussed.
\end{abstract}

\eqnobysec
\section{Introduction}

One of the central problems in computer science is to develop efficient
algorithms for hard optimization problems \cite{Garey-Johnson}.  A
standard approach is to propose a new algorithm for a given specific
problem by improving existing methods or by devising new approaches.
Simulated annealing (SA) presents a different perspective, which
provides a {\em generic} algorithm to be applicable in principle to an
arbitrary problem \cite{KGV,AK}.
The basic idea is to numerically simulate a physical annealing process by
the introduction of a temperature variable under the identification of
the cost function to be minimized with the energy of the system.
One decreases the temperature from a very high initial value toward zero as the
simulation time proceeds with the hope to reach the optimal state
(ground state) at the end of the process.

The efficiency of SA is determined by the choice of the annealing
schedule, the rate of temperature decrease.  A very slow decrease would
certainly lead the system to the ground state because the system stays
close to equilibrium at each temperature.  However, such a slow process
is not very useful practically.  On the other hand, when the temperature
is decreased too quickly, the system may be trapped in a local minimum.
It is therefore important to establish criteria on how fast one can
decrease the temperature to reach the optimal state avoiding local minima.

A theorem by Geman and Geman \cite{GG} gives a generic answer to this
problem. Any system is guaranteed to converge to the optimal state
in the limit of infinite time if the temperature is decreased
in proportion to $N/\log t$ or slower, where $N$ is the system size
and $t$ denotes simulation steps.
This result is highly non-trivial since the system reaches
the equilibrium state (ground state) after a long non-equilibrium
process in which the temperature changes with time at a finite,
non-vanishing, rate.

Quantum annealing (QA) is a relatively new alternative to SA, which uses
quantum fluctuations, instead of thermal effects, to search the phase
space of the system for the optimal state \cite{AHS,FGSSD,TH,KN,K,DC}.  An
artificial term of kinetic energy of quantum nature is introduced, by
which the system moves around in the phase space.  The cost function is
regarded as the potential energy.  A slow decrease of the kinetic energy
is expected to bring the system towards the optimal state.  A related
method of quantum adiabatic evolution \cite{FGGS} is based on
essentially the same idea.

A remarkable fact is that QA has been found to be more effective in
solving optimization problems than SA in most cases numerically
investigated so far, including the ground state search of random spin
systems \cite{SMTC,MST,SPA,SO}, protein folding \cite{LB}, the
configuration of molecules in a Lennard-Jones cluster \cite{LB2},
travelling salesman problem \cite{MST2}, simple potentials
\cite{SST05,SST06} and a kinetically constrained system \cite{DCS}.  It
has also been observed experimentally that a QA-like process leads to
equilibrium more efficiently than a thermal process \cite{BBRA}.  In
contrast, in the instance of 3-SAT, a hard optimization problem, QA has
been found not to outperform SA \cite{BST}.  It is therefore a very
interesting problem to establish when and how QA converges to the ground
state, preferably with a comparison with SA in mind.

In the present paper we report on a solution of this problem by proving
several theorems which give sufficient conditions for convergence of QA.
In many numerical studies of QA, stochastic processes are used in the
forms of path-integral and Green's function Monte Carlo simulations
mainly due to difficulties in directly solving the Schr\"odinger equation
for large systems.
Our approach reflects such developments, and we derive convergence
conditions for Monte Carlo implementations of QA using the idea of
Geman and Geman for convergence conditions for SA.

This paper consists of five sections. Various definitions of an
inhomogeneous Markov chain are given in the next section. Convergence of
QA is proved  for the path-integral and the
Green function Monte Carlos in section 3 and section 4,
respectively. The last section is devoted to discussions.

\section{Ergodicity of inhomogeneous Markov chain}
Since we use the theory of stochastic processes, it is useful to recall
various definitions and theorems for inhomogeneous Markov processes
\cite{AK}.  We denote the space of discrete states by $\mathcal{S}$ and
assume that the size of $\mathcal{S}$ is finite.  A Monte Carlo step is
characterized by the {\it transition probability} from state $x (\in
\mathcal{S})$ to state $y (\in\mathcal{S})$ at time step $t$:
\begin{equation}
 \label{eq:G}
 G(y,x;t)= \cases{P(y,x)A(y,x;t) & ($x\neq y$)\\
 1-\sum_{z\in\mathcal{S}}P(z,x)A(z,x;t) & ($x=y$)},
\end{equation}
where $P(y,x)$ and $A(y,x;t)$ are called the {\it generation probability}
and the {\it acceptance probability}, respectively.
The former is the probability to generate the next candidate state $y$
from the present state $x$. We assume that this probability does not depend
on time and satisfies the following conditions: 
%
\begin{eqnarray}
 \forall x,y\in\mathcal{S} : P(y,x)=P(x,y)\geq 0 ,\\
 \forall x\in\mathcal{S} : P(x,x)=0 ,\\
 \forall x\in\mathcal{S} : \sum_{y\in\mathcal{S}}P(y,x)=1,\\ \fl
 \forall x,y\in\mathcal{S}, \exists n>0, \exists
  z_1,\cdots,z_{n-1}\in{\mathcal S}:
 \prod_{k=0}^{n-1}P(z_{k+1},z_k)>0, z_0=x, z_n=y.
\end{eqnarray}
The last condition represents irreducibility of $\mathcal{S}$, that is,
any state in $\mathcal{S}$ can be reached from any other state in
$\mathcal{S}$.

We define $\mathcal{S}_x$ as the neighbourhood of $x$, i.e., the set
of states that can be reached by a single step from $x$:
\begin{equation}
 \mathcal{S}_x = \{y\,|\,y\in\mathcal{S}, P(y,x)>0\}.
\end{equation}
The acceptance probability $A(y,x;t)$ is the probability to accept the
candidate $y$ generated from state $x$. The matrix $G(t)$, whose
$(y, x)$ component is given by \eref{eq:G}, $[G(t)]_{y,x}=G(y,x;t)$,
is called the {\it transition matrix}.

Let $\mathcal{P}$ denote the set of probability distributions on
$\mathcal{S}$.  We regard a probability distribution $p
(\in\mathcal{P})$ as the column vector with the component
$[p]_x=p(x)$. The probability distribution at time $t$, started from an
initial distribution $p_0 (\in\mathcal{P})$ at time $t_0$, is written as
\begin{equation}
 p(t,t_0)= G^{t,t_0} p_0\equiv G(t-1)G(t-2)\cdots G(t_0)p_0.
\end{equation}

A Markov chain is called {\em inhomogeneous} when the transition
probability depends on time.  In sections 3 and 4, we will prove that
inhomogeneous Markov chains associated with QA are ergodic under
appropriate conditions.  There are two kinds of ergodicity, weak and
strong.  {\it Weak ergodicity} means that the probability distribution
becomes independent of initial conditions after a sufficiently long
time:
\begin{equation}
 \forall t_0\geq 0: \lim_{t\rightarrow\infty} \sup \{
  \| p(t,t_0)-p'(t,t_0)\| \,|\, p_0, p'_0\in\mathcal{P}\}=0,
\end{equation}
where $p(t,t_0)$ and $p'(t,t_0)$ are the probability distributions with
different initial distributions $p_0$ and $p'_0$. The norm is defined by
\begin{equation}
 \| p \|=\sum_{x\in\mathcal{S}} |p(x)|.
\end{equation}
{\it Strong ergodicity} is the property that the probability
distribution converges to a unique distribution irrespective of initial
state:
\begin{equation}
 \label{eq:SE}
 \exists r\in\mathcal{P}, \forall t_0\geq 0:
  \lim_{t\rightarrow\infty} \sup \{
  \|p(t,t_0)-r\| \,|\, p_0\in\mathcal{P} \}=0.
\end{equation}

The following two theorems provide conditions for weak and strong
ergodicity of an inhomogeneous Markov chain \cite{AK}.

\begin{theorem}[Condition for weak ergodicity]
An inhomogeneous Markov chain is weakly ergodic if and only if there
exists a strictly increasing sequence of positive numbers $\{t_i\},
(i=0,1,2,\dots)$, such that
\begin{equation}
 \sum_{i=0}^{\infty}\left(1-\alpha(G^{t_{i+1},t_i})
 \right)\longrightarrow\infty,
\end{equation}
where $\alpha(G^{t_{i+1},t_i})$ is the {\it coefficient of ergodicity}
defined by
\begin{equation}
 \alpha(G^{t_{i+1},t_i})=1-\min\left\{ \sum_{z\in\mathcal{S}}
        \min\{G(z,x),G(z,y)\} \Big| x,y\in\mathcal{S} \right\}
\end{equation}
with the notation $G(z,x)=[G^{t_{i+1},t_i}]_{z,x}$.
\label{theorem:weak-e}
\end{theorem}

\begin{theorem}[Condition for strong ergodicity]
An inhomogeneous Markov chain is strongly ergodic if the following
three conditions hold:
\begin{enumerate}
 \item the Markov chain is weakly ergodic,
 \item for all $t$ there exists a stationary state $p_t\in\mathcal{P}$
       such that $p_t=G(t)p_t$,
 \item $p_t$ satisfies
       \begin{equation}
        \sum_{t=0}^{\infty}\|p_t-p_{t+1}\|<\infty.
       \end{equation}
\end{enumerate}
Moreover, if $\displaystyle p=\lim_{t\rightarrow\infty}p_t$, then $p$
is equal to the probability distribution $r$ in \eref{eq:SE}.
\label{theorem:strong-e}
\end{theorem}

\section{Quantum annealing with path-integral Monte Carlo method}
\subsection{Path-integral Monte Carlo method}
Let us first discuss convergence conditions for the implementation of
quantum annealing by the path-integral Monte Carlo (PIMC) method \cite{LB3}.
The basic idea of PIMC is to apply the Monte Carlo method to the
classical system obtained from the original quantum system by the
path-integral formula.
It is instructive to first consider the example of ground state search
of the Ising spin system as a typical combinatorial optimization problem.
The Ising system with generic interactions as discussed below covers
a wide range of problems in combinatorial optimization.
Examples include the ground-state search of spin glasses, travelling
salesman problem, neural networks and the satisfiability problem,
many of which have been treated in the Ising expression
in the literature mentioned in Introduction.

Quantum fluctuations are introduced by adding a transverse field
to the usual Ising spin system.
The Hamiltonian of the transverse-field Ising model (TFIM) thus
obtained is written as
\begin{equation}
 H(t)=-\sum_{\langle ij\rangle}J_{ij} \sigma_i^z \sigma_j^z 
 -\Gamma(t)\sum_{i=0}^{N}\sigma_i^x ,\label{eq:H_TFIM}
\end{equation}
where the $\sigma_i^\alpha$ ($\alpha=x,y,z$) are the Pauli matrices,
components of the spin $\frac{1}{2}$ operator at site $i$, and $J_{ij}$
denotes the coupling constant between sites $i$ and $j$.  There is no
restriction in the spatial dimensionality and the lattice structure.  It
is also to be noted that the existence of arbitrary many-body
interactions between $z$ components of Pauli matrix and longitudinal
random magnetic field $\sum h_i \sigma_i^z$, in addition to the above
Hamiltonian, would not change the following argument.

The first term of the right-hand side of \eref{eq:H_TFIM} is the cost
function (or potential) to be minimized. The transverse field
$\Gamma(t)$ represents the strength of kinetic energy of quantum nature,
which induces spin flips between up and down states measured in the $z$
direction.  In the QA, $\Gamma(t)$ is gradually reduced from a very
large (or infinitely large) initial value to zero as time proceeds.  By
starting from the trivial ground state of the initial system composed
only of the transverse-field term $-\Gamma (t)\sum_i \sigma_i^{x}$ and
following the time development of the system under a slow decrease of
the transverse field, one hopes to eventually reach the non-trivial
ground state of the original problem, $-\sum J_{ij}\sigma_i^z
\sigma_j^z$, when $\Gamma (t)$ vanishes.  An important problem is how
slow is sufficiently slow to achieve this goal.

In the path-integral method, the $d$-dimensional TFIM is mapped to a
$(d+1)$-dimensional classical Ising system so that the quantum system
can be simulated on classical computer.  In numerical simulations, the
Suzuki-Trotter formula \cite{T,S} is usually employed to express the
partition function of the resulting classical system,
\begin{equation}\fl
 \label{eq:ST}
 Z(t)\approx \sum_{\{S_i^{(k)}\}} \exp\left(\frac{\beta}{M}\sum_{k=1}^M
 \sum_{\langle ij\rangle}J_{ij}S_i^{(k)}S_j^{(k)}
 +\gamma(t)\sum_{k=1}^{M}\sum_{i=0}^{N}S_i^{(k)}S_i^{(k+1)}\right),
\end{equation}
where $M$ is the length along the extra dimension (Trotter number) and
$S_i^{(k)} (=\pm 1)$ denotes a classical Ising spin at site $i$ on
the $k$th Trotter slice.
The nearest-neighbour interaction between adjacent Trotter slices,
\begin{equation}
 \label{eq:gamma}
 \gamma(t)=\frac{1}{2}\log\left(\coth\frac{\beta\Gamma(t)}{M}\right),
\end{equation}
is ferromagnetic.  This approximation \eref{eq:ST} becomes exact in the
limit $M\to\infty$ for a fixed $\beta=1/k_BT$.  The magnitude of this
interaction \eref{eq:gamma} increases with time $t$ and tends to
infinity as $t\rightarrow\infty$, reflecting the decrease of $\Gamma
(t)$.  We fix $M$ and $\beta$ to arbitrary large values, which
corresponds to the actual situation in numerical simulations.  Therefore
the theorem presented below does not directly guarantee the convergence
of the system to the true ground state, which is realized only after
taking the limits $M\to\infty$ and $\beta\to \infty$.  We will rather
show that the system converges to the thermal equilibrium represented by
the right-hand side of \eref{eq:ST}, which can be chosen arbitrarily
close to the true ground state by taking $M$ and $\beta$ large enough.

With the above example of TFIM in mind, it will be convenient to treat a
more general expression than \eref{eq:ST},
\begin{equation}
 \label{eq:Z_PIMC}
 Z(t)=\sum_{x\in\mathcal{S}} \exp
 \left(-\frac{F_0(x)}{T_0}-\frac{F_1(x)}{T_1(t)}\right).
\end{equation}
Here $F_0(x)$ is the cost function whose global minimum is the desired
solution of the combinatorial optimization problem. The temperature
$T_0$ is chosen to be sufficiently small.  The term $F_1(x)$ derives
from the kinetic energy, which is the transverse field in the TFIM.
Quantum fluctuations are tuned by the extra temperature factor $T_1(t)$,
which decreases with time.  The first term $-F_0(x)/T_0$ corresponds to
the interaction term in the exponent of \eref{eq:ST}, and the second
term $-F_1(x)/T_1(t)$ generalizes the transverse-field term in
\eref{eq:ST}.

For the partition function \eref{eq:Z_PIMC}, we define
the acceptance probability of PIMC as
\begin{eqnarray}
 \label{eq:A_PIMC}
 A(y,x;t)=g\left(\frac{q(y;t)}{q(x;t)}\right), \\
 \label{eq:BD}
 q(x;t)=\frac{1}{Z(t)}
 \exp\left(-\frac{F_0(x)}{T_0}-\frac{F_1(x)}{T_1(t)}\right).
\end{eqnarray}
This $q(x;t)$ is the equilibrium Boltzmann factor at a given fixed $T_1(t)$.
The function $g(u)$ is the {\it acceptance function}, a monotone increasing
function satisfying $0\leq g(u)\leq 1$ and $g(1/u)=g(u)/u$ for $u\geq 0$.
For instance, for the heat bath and the Metropolis methods, we have
\begin{eqnarray}
 g(u)=\frac{u}{1+u}, \\
 g(u)=\min\{1,u\},
\end{eqnarray}
respectively. The conditions mentioned above for $g(u)$ guarantee that
$q(x;t)$ is the stationary distribution of the homogeneous Markov chain
defined by the transition matrix $G(t)$ with a fixed $t$.  In other words,
$q(x;t)$ is the right eigenvector of $G(t)$ with eigenvalue 1.

\subsection{Convergence theorem for QA-PIMC}

We first define a few quantities.
The set of local maximum states of $F_1$ is written as $\mathcal{S}_m$,
\begin{equation}
 \mathcal{S}_m=\left\{x\,|\,x\in\mathcal{S},\ \forall y\in \mathcal{S}_x,\
  F_1(y)\leq F_1(x)\right\}.
\end{equation}
We denote by $d(y,x)$ the minimum number of steps
necessary to make a transition from $x$ to $y$. 
Using this notation we define the minimum number of maximum steps
needed to reach any other state from an arbitrary state in the set 
$\mathcal{S}\setminus\mathcal{S}_m$,
\begin{equation}
 R=\min\Bigl\{\max\left\{d(y,x)\,|\,y\in\mathcal{S}\right\}
 \Bigm| x\in\mathcal{S}\setminus\mathcal{S}_m \Bigr\}.
 \label{def:R}
\end{equation}
Also, $L_0$ and $L_1$ stand for the maximum changes of $F_0(x)$ and $F_1(x)$,
respectively, in a single step,
\begin{eqnarray}
 L_0=\max\left\{\left|F_0(x)-F_0(y)\right|\, |\, P(y,x)>0,\ 
 x,y\in\mathcal{S}\right\}, \\
 L_1=\max\left\{\left|F_1(x)-F_1(y)\right|\, |\, P(y,x)>0,\ 
 x,y\in\mathcal{S}\right\}.
\end{eqnarray}
Our main results are summarized in the following theorem and its corollary.
\begin{theorem}[Strong ergodicity of the system \eref{eq:Z_PIMC}]
The inhomogeneous Markov chain generated by \eref{eq:A_PIMC} and
\eref{eq:BD} is strongly ergodic and converges to the equilibrium state
corresponding to the first term of the right-hand side of \eref{eq:BD},
$\exp(-F_0(x)/T_0)$, if
\begin{equation}
 \label{eq:AS_PIMC}
 T_1(t)\geq \frac{RL_1}{\log(t+2)}.
\end{equation}
\label{theorem:PIMC}
\end{theorem}
Application of this theorem to the PIMC implementation of QA represented by
\eref{eq:ST} immediately yields the following corollary.
\begin{corollary}[Strong ergodicity of QA-PIMC for TFIM]
The inhomogeneous Markov chain generated by the Boltzmann factor on the
right-hand side of \eref{eq:ST} is strongly ergodic and converges to the
equilibrium state corresponding to the first term on the right-hand side
of \eref{eq:ST} if
\begin{equation}
 \Gamma(t)\geq\frac{M}{\beta}\tanh^{-1}\frac{1}{(t+2)^{2/RL_1}}.
\end{equation}
\label{corollary:PIMC}
\end{corollary}
\subsubsection*{Remark.} 
For sufficiently large $t$, the above inequality reduces to
\begin{equation}
 \Gamma(t)\geq \frac{M}{\beta}(t+2)^{-2/RL_1}.  \label{eq:power-decay}
\end{equation}
This result implies that a power decay of the transverse field is
sufficient to guarantee the convergence of quantum annealing of TFIM by
the PIMC.

To prove strong ergodicity it is necessary to prove weak ergodicity
first.  The following lemma is useful for this purpose.  The proof of
this lemma is given in \ref{appendix:A}.

\begin{lemma}[Lower bound on the transition probability]
The elements of the transition matrix defined by \eref{eq:G},
\eref{eq:A_PIMC} and \eref{eq:BD} have the following lower bound:
\begin{equation}\fl
 \label{eq:LB1}
 P(y,x)>0 \Rightarrow \forall t>0 : 
 G(y,x;t)\geq w\, g(1) \exp\left(-\frac{L_0}{T_0}-\frac{L_1}{T_1(t)}\right),
\end{equation} 
and
\begin{equation}\fl
 \label{eq:LB2}
 \exists t_1>0, \forall x\in\mathcal{S}\setminus\mathcal{S}_m, 
 \forall t\geq t_1
  : G(x,x;t)\geq w\, g(1) 
 \exp\left(-\frac{L_0}{T_0}-\frac{L_1}{T_1(t)}\right).
\end{equation}
\label{lemma:1}
\end{lemma}
\subsubsection*{Proof of weak ergodicity implied in Theorem \ref{theorem:PIMC}.} 
Let us introduce the following quantity
\begin{equation}
 x^*=\arg \min\Bigl\{\max\left\{d(y,x)\,|\,y\in\mathcal{S}\right\}
 \Bigm| x\in\mathcal{S}\setminus\mathcal{S}_m \Bigr\}.
\end{equation}
Comparison with the definition of $R$ in \eref{def:R} implies that
the state $x^*$ is reachable by at most $R$ transitions from any states.
Also, $w$ stands for the minimum non-vanishing value of $P(y,x)$,
\begin{equation}
 w = \min\left\{P(y,x)\,|\, P(y,x)>0,\ x,y\in\mathcal{S}\right\}.
\end{equation}

Now, consider the transition probability from an arbitrary state $x$ to
$x^*$. From the definitions of $R$ and $x^*$, there exists at least one
transition route within $R$ steps:
\begin{equation*}
 x\equiv x_0\neq x_1\neq x_2\neq \cdots \neq x_l
 =x_{l+1}=\cdots =x_R\equiv x^* .
\end{equation*}
Then Lemma \ref{lemma:1} yields that, for sufficiently large $t$, the
transition probability at each time step has the following lower bound:
\begin{equation}
 G(x_{i+1},x_i;t-R+i)\geq w g(1) 
 \exp\left(-\frac{L_0}{T_0}-\frac{L_1}{T_1(t-R+i)}\right).
  \label{eq:G-bound1}
\end{equation}
Thus, by taking the product of \eref{eq:G-bound1} from $i=0$ to $i=R-1$,
we have
\begin{eqnarray}\fl
 G^{t,t-R}(x^*,x)&\geq
 G(x^*,x_{R-1};t-1)G(x_{R-1},x_{R-2};t-2)\cdots G(x_1,x;t-R)
 \nonumber \\ \fl
 &\geq \prod_{i=0}^{R-1} w\, g(1)
 \exp\left(-\frac{L_0}{T_0}-\frac{L_1}{T_1(t-R+i)}\right)
 \nonumber \\ \fl
 &\geq w^R g(1)^R
 \exp\left(-\frac{RL_0}{T_0}-\frac{RL_1}{T_1(t-1)}\right),
  \label{eq:G-bound2}
\end{eqnarray}
where we have used monotonicity of $T_1(t)$.
Consequently, it is possible to find an integer $k_0\geq 0$ such that,
for all $k>k_0$, the coefficient of ergodicity satisfies
\begin{equation}
 1-\alpha(G^{kR,kR-R}))\geq w^R g(1)^R 
 \exp\left(-\frac{RL_0}{T_0}-\frac{RL_1}{T_1(kR-1)}\right).
 \label{eq:1-minus-alpha}
\end{equation}
We now substitute the annealing schedule \eref{eq:AS_PIMC}.  Then weak
ergodicity is immediately proved from Theorem \ref{theorem:weak-e}
because we obtain
\begin{equation}\fl
 \sum_{k=1}^{\infty} (1-\alpha(G^{kR,kR-R}))
 \geq w^R g(1)^R \exp\left(-\frac{RL_0}{T_0}\right)
 \sum_{k=k_0}^{\infty}\frac{1}{kR+1}\longrightarrow\infty.
\end{equation}
\subsubsection*{Proof of Theorem \ref{theorem:PIMC}.}
To prove strong ergodicity, we refer to Theorem \ref{theorem:strong-e}.
The condition (i) has already been proved.  As has been mentioned, the
Boltzmann factor \eref{eq:BD} satisfies $q(t)=G(t)q(t)$, which is the
condition (ii).  Thus the proof will be complete if we prove the
condition (iii) by setting $p_t=q(t)$.  As shown in \ref{appendix:B},
$q(x;t)$ is monotonically increasing for large $t$:
\begin{equation}
 \label{eq:MBD1}
 \forall t\geq 0, \forall x\in\mathcal{S}_1^{\rm min} :
 q(x;t+1)\geq q(x;t),
\end{equation}
\begin{equation}
 \label{eq:MBD2}
 \exists t_1>0, \forall t\geq t_1, \forall
  x\in\mathcal{S}\setminus\mathcal{S}_1^{\rm min} :
 q(x;t+1)\leq q(x;t),
\end{equation}
where $\mathcal{S}_1^{\rm min}$ denotes the set of global minimum states
of $F_1$. Consequently, for all $t>t_1$, we have
\begin{eqnarray}\fl
 \|q(t+1)-q(t)\| 
 &=\sum_{x\in\mathcal{S}_1^{\rm min}}\left\{
 q(x;t+1)-q(x;t)\right\}
 -\sum_{x\not\in\mathcal{S}_1^{\rm min}}\left\{
 q(x;t+1)-q(x;t)\right\} \nonumber \\
 &=2\sum_{x\in\mathcal{S}_1^{\rm min}}\left\{
 q(x;t+1)-q(x;t)\right\},
 \label{eq:qq-difference}
\end{eqnarray}
where we used $\|q(t)\|=\sum_{x\in\mathcal{S}_1^{\rm min}}q(x;t)
+\sum_{x\not\in\mathcal{S}_1^{\rm min}}q(x;t)=1$. We then obtain
\begin{equation}
 \sum_{t=t_1}^{\infty}\|q(t+1)-q(t)\|
 =2\sum_{x\in\mathcal{S}_1^{\rm min}}\left\{
 q(x;\infty)-q(x;t_1)\right\}\leq 2.
\end{equation}
Therefore $q(t)$ satisfies the condition (iii):
\begin{eqnarray}
 \sum_{t=0}^{\infty}\|q(t+1)-q(t)\| &=
 \sum_{t=0}^{t_1-1}\|q(t+1)-q(t)\|+\sum_{t=t_1}^{\infty}
 \|q(t+1)-q(t)\| \nonumber \\
 &\leq 2 t_1+2<\infty,
\end{eqnarray}
which completes the proof of strong ergodicity.

\subsection{Remarks}

\subsubsection*{Remark 1.}
In the above analyses we treated systems with discrete degrees of
freedom.  Theorem \ref{theorem:PIMC} does not apply directly to a
continuous system.  Nevertheless, by discretization of the continuous
space we obtain the following result.

Let us consider a system of $N$ distinguishable particles in a continuous
space of finite volume with the Hamiltonian
\begin{equation}
 \label{eq:H}
 H=\frac{1}{2m(t)}\sum_{i=1}^{N}\bi{p}_i^2+V(\{\bi{r}_i\}).
\end{equation}
The mass $m(t)$ controls the magnitude of quantum fluctuations.  The
goal is to find the minimum of the potential term, which is achieved by
a gradual increase of $m(t)$ to infinity according to the prescription
of QA.
After discretization of the continuous space (which is necessary anyway
in any computer simulations with finite precision) and an application of
the Suzuki-Trotter formula, the equilibrium partition function acquires
the following expression in the representation to diagonalize spatial
coordinates
\begin{equation}\fl
  Z(t)\approx \Tr\exp\left(
 -\frac{\beta}{M}\sum_{k=1}^{M}V\left(\{\bi{r}_i^{(k)}\}\right)
 -\frac{Mm(t)}{2\beta}\sum_{i=1}^{N}\sum_{k=1}^{M}
  \left|\bi{r}_i^{(k+1)}-\bi{r}_i^{(k)}\right|^2 \right),
\end{equation}
where we choose the unit $\hbar=1$.  Theorem \ref{theorem:PIMC} is
applicable to this system under the identification of $T_1(t)$ with
$m(t)^{-1}$.  We therefore conclude that a logarithmic increase of the
mass suffices to guarantee strong ergodicity of the
potential-minimization problem under spatial discretization.

The coefficient corresponding to the numerator of the right-hand side of
\eref{eq:AS_PIMC} is estimated as
\begin{equation}
 RL_1 \approx M^2NL^2/\beta,
 \label{eq:RL_PIMC}
\end{equation}
where $L$ denotes the maximum value of
$\left|\bi{r}_i^{(k+1)}-\bi{r}_i^{(k)}\right|$.  To obtain this
coefficient, let us consider two extremes. One is that any states
are reachable at one step. By definition, $R=1$ and $L_1\approx M^2 N
L^2/\beta$, which yield \eref{eq:RL_PIMC}. The other case is that only
one particle can move to the nearest neighbour point at one time
step. With $a$ $(\ll L)$ denoting the lattice spacing, we have
\begin{equation}
 L_1 \approx \frac{M}{2\beta}\left\{L^2-(L-a)^2\right\}
 \approx\frac{ML a}{\beta}.
\end{equation}
Since the number of steps to reach any configurations is
estimated as $R\approx NML/a$, we again obtain \eref{eq:RL_PIMC}.

\subsubsection*{Remark 2.}
In Theorem \ref{theorem:PIMC}, the acceptance probability is defined by
the conventional Boltzmann form, \eref{eq:A_PIMC} and \eref{eq:BD}.
However, we have the freedom to choose any transition (acceptance)
probability as long as it is useful to achieve our objective since our
goal is not to find finite-temperature equilibrium states but to
identify the optimal state.
There have been attempts to accelerate the annealing schedule in SA by
modifying the transition probability.  In particular Nishimori and Inoue
\cite{NI} have proved weak ergodicity of the inhomogeneous Markov chain
for classical simulated annealing using the probability of Tsallis and
Stariolo \cite{TS}.  There the property of weak ergodicity was shown to
hold under the annealing schedule of temperature inversely proportional
to a power of time steps.  This annealing rate is much faster than the
log-inverse law of Geman and Geman for the conventional Boltzmann
factor.

A similar generalization is possible for QA-PIMC by using 
the following modified acceptance probability
\begin{eqnarray}
 A(y,x;t)=g\left(u(y,x;t)\right), \\ \fl
 u(y,x;t)=\e^{-({F_0(y)-F_0(x)})/{T_0}}
 \left\{1+(q-1)\frac{F_1(y)-F_1(x)}{T_1(t)}\right\}^{1/(1-q)},
\end{eqnarray}
where $q$ is a real number.
In the limit $q\rightarrow 1$, this acceptance probability reduces to the
Boltzmann form.
Similarly to the discussions leading to Theorem \ref{theorem:PIMC},
we can prove that the inhomogeneous Markov chain with this acceptance
probability is weakly ergodic if
\begin{equation}
 T_1(t)\geq \frac{b}{(t+2)^c}, \qquad 0< c\leq \frac{q-1}{R},
 \label{eq:T_G_PIMC}
\end{equation}
where $b$ is a positive constant.  We have to restrict ourselves to the
case $q>1$ for a technical reason as was the case previously \cite{NI}.
We do not reproduce the proof here because it is quite straightforward
to generalize the discussions for Theorem \ref{theorem:PIMC} in
combination with the argument of \cite{NI}.
The result \eref{eq:T_G_PIMC} applied to the TFIM is that, if the
annealing schedule asymptotically satisfies
\begin{equation}
 \Gamma(t)\geq \frac{M}{\beta}\exp\left(-\frac{2(t+2)^c}{b}\right),
\end{equation}
the inhomogeneous Markov chain is weakly ergodic. Notice that
this annealing schedule is faster than the power law of \eref{eq:power-decay}.
We have been unable to prove strong ergodicity because we could not
identify the stationary distribution for a fixed $T_1(t)$ in the present case.

\section{Quantum annealing with Green's function Monte Carlo method}

The path-integral Monte Carlo simulates only the equilibrium behaviour
at finite temperature because its starting point is the equilibrium
partition function.  Moreover, it follows an artificial time evolution
of Monte Carlo dynamics, not the natural Schr\"odinger dynamics.  An
alternative approach to improve these points is the Green's function
Monte Carlo (GFMC) method \cite{LB3,CA,TC}.  The basic idea is to solve
the imaginary-time Schr\"{o}dinger equation by stochastic processes.
The Schr\"{o}dinger dynamics with imaginary time has an extra advantage
that one can reach the optimal state more efficiently than by real-time
dynamics \cite{SST}.  Thus, for our purpose to solve optimization
problems, it is more important to discuss imaginary-time Schr\"{o}dinger
equation than the ``natural'' real-time evolution.

In the present section we derive sufficient conditions for strong
ergodicity to hold in GFMC.

\subsection{Green's function Monte Carlo method}

The evolution of states by the imaginary-time Schr\"{o}dinger equation
starting from an initial state $|\psi_0\rangle$ is expressed as
\begin{equation}
  |\psi(t)\rangle = {\rm T}\exp\left(-\int_{0}^{t}\rmd t' H(t')\right)
   |\psi_0\rangle,
\end{equation}
where T is the time-ordering operator.  The right-hand side can be
decomposed into a product of small-time evolutions,
\begin{equation}
 |\psi(t)\rangle =\lim_{n\rightarrow\infty}
  \hat{G}_0(t_{n-1})\hat{G}_0(t_{n-2})
  \cdots\hat{G}_0(t_1)\hat{G}_0(t_0)|\psi_0\rangle,
  \label{eq:GFMC1}
\end{equation}
where $t_k=k\Delta t$, $\Delta t=t/n$ and $\hat{G}_0(t)=1-\Delta t\cdot
H(t)$.  In the GFMC, one approximates the right-hand side of this
equation by a product with large but finite $n$ and replaces
$\hat{G}_0(t)$ with $\hat{G}_1(t)=1-\Delta t (H(t)-E_T)$, where $E_T$ is
called the reference energy to be taken approximately close to the final
ground-state energy.  This subtraction of the reference energy simply
adjusts the standard of energy and changes nothing physically.  However,
practically, this term is important to keep the matrix elements positive
and to accelerate convergence to the ground state as will be explained
shortly.

To realize the process of \eref{eq:GFMC1} by a stochastic method,
we rewrite this equation in a recursive form,
\begin{equation}
 \label{eq:psi}
 \psi_{k+1}(y) = \sum_{x} \hat{G}_1(y,x;t_k)\psi_k(x),
\end{equation}
where $\psi_k(x)=\langle x|\psi_k\rangle$ and $|x\rangle$ denotes a
basis state.  The matrix element of Green's function is given by
\begin{equation}
 \label{eq:G1_GFMC}
 \hat{G}_1(y,x;t)=\langle y|1-\Delta t(H(t)-E_T)|x\rangle.
\end{equation}
Equation \eref{eq:psi} looks similar to a Markov process
but is significantly different in several ways.
An important difference is that the Green's function is not normalized,
$\sum_{y}\hat{G}_1(y,x;t)\neq 1$.
In order to avoid this problem, one decomposes the Green's function into a
normalized probability $G_1$ and a weight $w$:
\begin{equation}
  \hat{G}_1(y,x;t)=G_1(y,x;t) w(x;t),
\end{equation}
where
\begin{equation}
  G_1(y,x;t)\equiv \frac{\hat{G}_1(y,x;t)}{\sum_y \hat{G}_1(y,x;t)}, \quad
  w(x;t)\equiv \frac{\hat{G}_1(y,x;t)}{G_1(y,x;t)}.
\end{equation}
Thus, using \eref{eq:psi}, the wave function at time $t$ is written as
\begin{eqnarray}\fl 
 \label{eq:GFMC}
 \psi_n(y) &=\sum_{\{x_k\}}\delta_{y,x_n}w(x_{n-1};t_{n-1})
 w(x_{n-2};t_{n-2})\cdots w(x_0;t_0) \nonumber \\ \fl
 &\quad\times G_1(x_n,x_{n-1};t_{n-1}) G_1(x_{n-1},x_{n-2};t_{n-2})
 \cdots G_1(x_1,x_0;t_0) \psi_0(x_0).
\end{eqnarray}

The algorithm of GFMC is based on this formula and is defined by a
weighted random walk in the following sense.  One first prepares an
arbitrary initial wave function $\psi_0(x_0)$, all elements of which are
non-negative.  A random walker is generated, which sits initially
($t=t_0$) at the position $x_0$ with a probability proportional to
$\psi_0(x_0)$.  Then the walker moves to a new position $x_1$ following
the transition probability $G_1(x_1,x_0;t_0)$.  Thus this probability
should be chosen non-negative by choosing parameters appropriately as
described later.
Simultaneously, the weight of this walker is updated by the rule
$W_{1}=w(x_{0};t_{0})W_0$ with $W_0=1$.
This stochastic process is repeated to $t=t_{n-1}$.
One actually prepares $M$ independent walkers and let those walkers
follow the above process.
Then, according to \eref{eq:GFMC}, the wave function $\psi_n(y)$
is approximated by the distribution of walkers at the final step
weighted by $W_n$,
\begin{equation}
 \psi_n(y)= \lim_{M\rightarrow\infty}\frac{1}{M}
  \sum_{i=1}^{M}W_n^{(i)}\delta_{y,x_n^{(i)}},
\end{equation}
where $i$ is the index of a walker.

As noted above, $G_1(y,x;t)$ should be non-negative, which is achieved by
choosing sufficiently small $\Delta t$ (i.e. sufficiently large $n$)
and selecting $E_T$ within the instantaneous spectrum of the
Hamiltonian $H(t)$.
In particular, when $E_T$ is close to the instantaneous ground-state energy
of $H(t)$ for large $t$ (i.e. the final target energy), $\hat{G}_1(x,x;t)$
is close to unity whereas other matrix components of $\hat{G}_1(t)$
are small.
Thus, by choosing $E_T$ this way, one can accelerate convergence of GFMC
to the optimal state in the last steps of the process.

If we apply this general framework to the TFIM with the
$\sigma^z$-diagonal basis, the matrix elements of Green's function are
immediately calculated as
\begin{equation}\fl
  \hat{G}_1(y,x;t)=\cases{
  1-\Delta t(E_0(x)-E_T) & $(x=y)$ \\
  \Delta t\, \Gamma(t) & ($x$ and $y$ differ by a single-spin flip) \\
  0 & (otherwise),
  }
  \label{eq:g-hat}
\end{equation}
where $E_0(x)=\langle x| \left(-\sum_{ij} J_{ij}\sigma_i^z\sigma_j^z
\right) |x\rangle$.
One should choose $\Delta t$ and $E_T$ such that
$1-\Delta t(E_0(x)-E_T)\geq 0$ for all $x$.
Since $w(x,t)=\sum_{y}\hat{G}_1(y,x;t)$, the weight is given by
\begin{equation}
 \label{eq:w-GFMC}
 w(x;t)=1-\Delta t(E_0(x)-E_T)+N\Delta t\, \Gamma(t).
\end{equation}
One can decompose this transition probability into the generation probability
and the acceptance probability as in \eref{eq:G}:
\begin{equation}
 \label{eq:P_GFMC}
 P(y,x)=\cases{
  \frac{1}{N} & (single-spin flip) \\
  0 & (otherwise)}
\end{equation}
\begin{equation}
 \label{eq:A_GFMC}
 A(y,x;t)=\frac{N\Delta t\, \Gamma(t)}{1-\Delta t(E_0(x)-E_T)+N\Delta
  t\, \Gamma(t)}.
\end{equation}
We shall analyze the convergence properties of stochastic processes
under these probabilities for TFIM.

\subsection{Convergence theorem for QA-GFMC}

Similarly to the QA by PIMC, it is necessary to reduce the strength of
quantum fluctuations slowly enough in order to find the ground state in the GFMC.
The following theorem provides a sufficient condition in this regard.
\begin{theorem}[Strong ergodicity of QA-GFMC]
\label{theorem:GFMC}
The inhomogeneous Markov process of random walker for the QA-GFMC of TFIM,
\eref{eq:G}, \eref{eq:P_GFMC} and \eref{eq:A_GFMC}, is strongly ergodic if
\begin{equation}
 \label{eq:AS_GFMC}
 \Gamma(t)\geq \frac{b}{(t+1)^c}, \qquad 0< c\leq \frac{1}{N}.
\end{equation}
\end{theorem}

The lower bound of the transition probability given in the following lemma
will be used in the proof of Theorem \ref{theorem:GFMC}.

\begin{lemma}
\label{lemma:GFMC}
The transition probability of random walk in the GFMC defined by \eref{eq:G},
\eref{eq:P_GFMC} and \eref{eq:A_GFMC} has the lower bound:
\begin{equation}\fl
 P(y,x)>0 \Rightarrow \forall t>0: G_1(y,x;t)\geq
 \frac{\Delta t\,\Gamma(t)}{1-\Delta t(E_{\rm min}-E_T)
 +N\Delta t\,\Gamma(t)},
\end{equation}
\begin{equation}\fl
 \label{eq:L2_2}
 \exists t_1>0, \forall t>t_1: G_1(x,x;t)\geq
 \frac{\Delta t\,\Gamma(t)}{1-\Delta t(E_{\rm min}-E_T)
 +N\Delta t\,\Gamma(t)},
\end{equation}
where $E_{\rm min}$ is the minimum value of $E_0(x)$
\begin{equation}
 E_{\rm min}=\min\{ E_0(x)|x\in \mathcal{S}\}.
\end{equation}
\end{lemma}

\subsubsection*{Proof of Lemma \ref{lemma:GFMC}.}
The first part of Lemma \ref{lemma:GFMC} is trivial because the
transition probability is an increasing function with respect to
$E_0(x)$ when $P(y,x)>0$ as seen in \eref{eq:A_GFMC}.  Next, we prove
the second part of Lemma \ref{lemma:GFMC}.  According to \eref{eq:g-hat}
and \eref{eq:w-GFMC}, $G_1(x,x;t)$ is written as
\begin{equation}
 G_1(x,x;t)=1-\frac{N\Delta t\, \Gamma(t)}{1-\Delta t(E_0(x)-E_T)
 +N\Delta t\,\Gamma(t)}.
\end{equation}
Since the transverse field $\Gamma(t)$ decreases to zero with time, the
second term on the right-hand side tends to zero as $t \rightarrow
\infty$. Thus, there exists $t_1>0$ such that $G_1(x,x;t)>1-\varepsilon$
for $\forall \varepsilon>0$ and $\forall t>t_1$. On the other hand, the
right-hand side of \eref{eq:L2_2} converges to zero as
$t\rightarrow\infty$. We therefore have \eref{eq:L2_2}.

\subsubsection*{Proof of Theorem \ref{theorem:GFMC}.}
We show that the condition \eref{eq:AS_GFMC} is sufficient to
satisfy the three conditions of Theorem \ref{theorem:strong-e}.

(i) From Lemma \ref{lemma:GFMC}, we obtain a bound on the coefficient of
ergodicity for sufficiently large $k$ as
\begin{equation}\fl
  1-\alpha(G_1^{kN,kN-N})\geq \left\{
  \frac{\Delta t\, \Gamma(kN-1)}
  {1-\Delta t(E_{\rm min}-E_T)+N\Delta t\,\Gamma(kN-1)}\right\}^N,
\end{equation}
in the same manner as we derived \eref{eq:1-minus-alpha}, where we used $R=N$.
Substituting the annealing schedule \eref{eq:AS_GFMC}, we can prove weak
ergodicity from Theorem \ref{theorem:weak-e} because
\begin{equation}
 \sum_{k=1}^{\infty}\left(1-\alpha(G_1^{kN,kN-N})\right)\geq
 \sum_{k=k_0}^{\infty}\frac{b^N}{(kN)^{cN}}
\end{equation}
which diverges when $0<c\leq 1/N$.

(ii) As shown in \ref{appendix:C}, the stationary distribution of the
instantaneous transition probability $G_1(y,x;t)$ is
\begin{equation}
 \label{eq:q_GFMC}
 q(x;t)\equiv \frac{w(x;t)}{\sum_{x\in{\mathcal S}}w(x;t)}
 =\frac{1}{2^N}-\frac{\Delta t\, E_0(x)}
 {2^N \left\{1+\Delta t\, E_T+N\Delta t\, \Gamma(t)\right\}}.
\end{equation}

(iii) Since the transverse field $\Gamma(t)$ decreases monotonically
with $t$, the above stationary distribution $q(x;t)$ is an increasing
function of $t$ if $E_0(x)< 0$ and is decreasing if $E_0\geq 0$.
Consequently, using the same procedure as in \eref{eq:qq-difference}, we
have
\begin{equation}
 \|q(t+1)-q(t)\|=2\sum_{E_0(x)<0} \{q(x;t+1)-q(x;t)\},
\end{equation}
and thus
\begin{equation}
 \sum_{t=0}^{\infty}\|q(t+1)-q(t)\|=2\sum_{E_0(x)<0}\{q(x;\infty)-q(x;0)\}
 \leq 2.
\end{equation}
Therefore the sum $\sum_{t=0}^{\infty}\|q(t+1)-q(t)\|$ is finite, which
completes the proof of the condition (iii). 
\subsubsection*{Remark.}
Theorem \ref{theorem:GFMC} asserts convergence of the distribution of
random walkers to the equilibrium distribution \eref{eq:q_GFMC} with
$\Gamma(t)\to 0$.  This implies that the final distribution is not
delta-peaked at the ground state with minimum $E_0(x)$ but is a
relatively mild function of this energy.  The optimality of the solution
is achieved after one takes the weight factor $w(x;t)$ into account: The
repeated multiplication of weight factors as in \eref{eq:GFMC}, in
conjunction with the relatively mild distribution coming from the
product of $G_1$ as mentioned above, leads to the asymptotically
delta-peaked wave function $\psi_n(y)$ because $w(x;t)$ is larger for
smaller $E_0(x)$ as seen in \eref{eq:w-GFMC}.

\subsection{Alternative choice of Green's function}
So far we have used the Green's function defined in \eref{eq:G1_GFMC},
which is linear in the transverse field, allowing single-spin flips
only.  It may be useful to consider another type of Green's function
which accommodates multi-spin flips.  Let us try the following form of
Green's function,
\begin{equation}
  \label{eq:Green2}
  \hat{G}_2(t)=\exp\left(\Delta t\,\Gamma(t)\sum_{i}\sigma_i^x\right)
  \exp\left(\Delta t\sum_{ij}J_{ij}\sigma_i^z\sigma_j^z\right),
\end{equation}
which is equal to $\hat{G}_0(t)$ to the order $\Delta t$.
The matrix element of $\hat{G}_2(t)$ in the $\sigma^z$-diagonal basis is
\begin{equation}
  \hat{G}_2(y,x;t)=\cosh^N \left(\Delta t\,\Gamma(t)\right) 
  \tanh^\delta \left(\Delta t\,\Gamma(t)\right)\rme^{-\Delta t\,E_0(x)},
\end{equation}
where $\delta$ is the number of spins in different states in $x$ and $y$.
According to the scheme of GFMC, we decompose $\hat{G}_2(y,x;t)$
into the normalized transition probability and the weight:
\begin{equation}
 \label{eq:G2}
 G_2(y,x;t)=\left\{
 \frac{\cosh (\Delta t\,\Gamma(t))}{\rme^{\Delta t\,\Gamma(t)}}\right\}^N
 \tanh^\delta (\Delta t\,\Gamma(t)),
\end{equation}
\begin{equation}
 w_2(x;t)=\rme^{\Delta t\,N\Gamma(t)}\rme^{-\Delta t\,E_0(x)}.
\end{equation}
It is remarkable that the transition probability $G_2$ is independent of
$E_0(x)$. Thus, the stationary distribution of random walk is
uniform. This property is lost if one interchanges the
order of the two factors in \eref{eq:Green2}.

The property of strong ergodicity can be shown to hold in this case as well:
\begin{theorem}[Strong ergodicity of QA-GFMC 2]
The inhomogeneous Markov chain generated by \eref{eq:G2} is strongly
ergodic if
\begin{equation}
 \label{eq:AS2_GFMC}
\Gamma(t)\geq-\frac{1}{2\Delta t}\log\left\{
 1-2b(t+1)^{-1/N}\right\}.
\end{equation}
\end{theorem}

\subsubsection*{Remark.}
For sufficiently large $t$, the above annealing schedule is reduced to
\begin{equation}
 \Gamma(t)\geq \frac{b}{\Delta t\, (t+1)^{1/N}}.
\end{equation}

Since the proof is quite similar to the previous cases, we just outline the
idea of the proof.
The transition probability $G_2(y,x;t)$ becomes smallest when $\delta=N$.
Consequently, the coefficient of ergodicity is estimated as
\begin{equation*}
  1-\alpha(G_2^{t+1,t})\geq \left\{
  \frac{1-\rme^{-2\Delta t\,\Gamma(t)}}{2}\right\}^N.
\end{equation*}
We note that $R$ is equal to 1 in the present case because any states are
reachable from an arbitrary state in a single step.
From Theorem \ref{theorem:weak-e}, the condition
\begin{equation}
 \left\{\frac{1-\rme^{-2\Delta t\,\Gamma(t)}}{2}\right\}^N
 \geq \frac{b'}{t+1}
\end{equation}
is sufficient for weak ergodicity. From this, one obtains
\eref{eq:AS2_GFMC}.  Since the stationary distribution of $G_2(y,x;t)$
is uniform as mentioned above, strong ergodicity readily follows from
Theorem \ref{theorem:strong-e}.

Similarly to the case of PIMC, we can discuss the convergence condition of
QA-GFMC in systems with continuous degrees of freedom.
The resulting sufficient condition is a logarithmic
increase of the mass as will be shown now.
The operator $\hat{G}_2$ generated by the Hamiltonian \eref{eq:H} is
written as
\begin{equation}
  \hat{G}_2(t)=\exp\left(-\frac{\Delta t}{2m(t)}
  \sum_{i=1}^{N}\bi{p}_i^2\right)
  \rme^{-\Delta t V(\{\bi{r}_i\})}.
\end{equation} 
Thus, the Green's function is calculated in a discretized space as
\begin{equation}
 \hat{G}_2(y,x;t)\propto \exp\left(
 -\frac{m(t)}{2\Delta t}\sum_{i=1}^{N}\left|\bi{r}'_i-\bi{r}_i\right|^2
 -\Delta t V(\{\bi{r}_i\})\right),
\end{equation}
where $x$ and $y$ represent $\{\bi{r}_i\}$ and $\{\bi{r}'_i\}$, respectively.
Summation over $y$, i.e., integration over $\{\bi{r}'_i\}$, yields the weight
$w(x;t)$, from which the transition probability is obtained:
\begin{equation}
 w(x;t)\propto \rme^{-\Delta t V(\{\bi{r}_i\})},
\end{equation}
\begin{equation}
 G_2(y,x;t)\propto \exp\left( -\frac{m(t)}{2\Delta t}
 \sum_{i=1}^{N}\left|\bi{r}'_i-\bi{r}_i\right|^2\right).
\end{equation}
The lower bound for the transition probability depends exponentially on
the mass: $G_2(y,x;t)\geq \rme^{-Cm(t)}$. Since $1-\alpha(G_2^{t+1,t})$
has the same lower bound, the sufficient condition for weak ergodicity
is $\rme^{-Cm(t)}\geq (t+1)^{-1}$, which is rewritten as
\begin{equation}
 m(t)\leq C^{-1}\log(t+1).
\end{equation}
The constant $C$ is proportional to $NL^2/\Delta t$, where $L$ denotes
the maximum value of $|\bi{r}'-\bi{r}|$. The derivation of $C$ is
similar to \eref{eq:RL_PIMC}, because $G_2(t)$ allows any transition to
arbitrary states at one time step.

\section{Discussion}
We have proved strong ergodicity of the inhomogeneous Markov chains
associated with QA-PIMC and QA-GFMC, mainly with the application to
the TFIM in mind, which covers a wide range of combinatorial optimization
problems.
Our proof is quite general in the sense that it does not depend on
the spatial dimensionality or the lattice structure of the system.
The convergence of QA is guaranteed if the transverse field decreases as
$\Gamma(t) \approx {\rm const}/t^c$ asymptotically.
This annealing schedule for the
transverse field is faster than the temperature-annealing schedule,
the log-inverse law, found by Geman and Geman for SA.
Moreover, the generalized transition probability in PIMC accelerates
the annealing schedule to $\Gamma(t)\approx \exp (-t^c)$
(although we could not prove strong ergodicity in this case).
Since the constant $c$ appearing in these formulas
depends on the system size as $1/N$ and is therefore very small for large systems,
our result may not provide practically useful guidelines
to anneal the transverse field.
This is the same situation as in SA, in which the temperature annealing
should be $N/\log t$ or slower to converge.
Nevertheless our Theorems and Corollary represent quite non-trivial results
because they assure eventual convergence of the system to the ground
state (or a state near the ground state for PIMC) after
non-stationary processes without being trapped in local minima.

Let us write a few words on  computational complexity.
Although the annealing schedule of QA, the power-law dependence on $t$, is
much faster than the log-inverse law for SA, this does not mean that
QA provides an algorithm to solve NP problems in polynomial time.
The time for $\Gamma(t)$ to reach a sufficiently small value $\delta$
is estimated from \eref{eq:power-decay} as
\begin{equation}
 t_{1}\sim\exp\left(
 \frac{RL_1}{2}\log\frac{M}{\beta\delta}\right).
\end{equation}
Since $RL_1$ is of the order of $N$, the QA needs a time
exponential in $N$ to converge.
An important point is that the coefficient of $N$ in the exponent,
$\mathcal{O}(\log \delta^{-1} )$, is much smaller than that for SA,
in which the coefficient is $\mathcal{O}(1/\delta )$ as can be seen
from $T(t)\approx N/\log t \approx \delta$.
The situation is the same in QA-GFMC.
If one uses the generalized transition probability, the
corresponding time is
\begin{equation}
 t_{2}\sim\exp\left(N \log\left(\log\frac{1}{\delta}\right)\right),
\end{equation}
which again shows exponential dependence on $N$ with a much smaller
coefficient.

\ack

This work was partially supported by CREST, JST. One of the authors (S.M.)
is supported by Research Fellowships of the Japan Society for the
Promotion of Science for Young Scientists.

\appendix
\section{Proof of Lemma \ref{lemma:1}}\label{appendix:A}
The first part of Lemma \ref{lemma:1} is proved straightforwardly.
\Eref{eq:LB1} follows directly from the definition of the transition
probability and the property of the acceptance function $g$.
When $q(y;t)/q(x;t)<1$, we have
\begin{equation}\fl
  G(y,x;t)\geq w\, g\left(\frac{q(x;t)}{q(y;t)}\right)
 \frac{q(y;t)}{q(x;t)}\geq w\, g(1) 
 \exp\left(-\frac{L_0}{T_0}-\frac{L_1}{T_1(t)}\right).
\end{equation}
On the other hand, if $q(y;t)/q(x;t)\geq1$,
\begin{equation}
 G(y,x;t)\geq w\, g(1)\geq w\, g(1) 
 \exp\left(-\frac{L_0}{T_0}-\frac{L_1}{T_1(t)}\right),
\end{equation}
where we used the fact that both $L_0$ and $L_1$ are positive.

Next, we prove \eref{eq:LB2}. Since $x$ is not a member of $\mathcal{S}_m$,
there exists a state $y\in\mathcal{S}_x$ such that
$F_1(y)-F_1(x)>0$. For such a state $y$,
\begin{equation}
 \lim_{t\rightarrow\infty} g\left(\exp\left(
 -\frac{F_0(y)-F_0(x)}{T_0}-\frac{F_1(y)-F_1(x)}{T_1(t)}\right)\right)
 =0,
\end{equation}
because $T_1(t)$ tends to zero as $t\rightarrow\infty$ and $0\leq g(u)\leq
u$. Thus, for all $\varepsilon>0$, there exists $t_1>0$  such that
\begin{equation}
 \forall t>t_1 : g\left(\exp\left(-\frac{F_0(y)-F_0(x)}{T_0}
 -\frac{F_1(y)-F_1(x)}{T_1(t)}\right)\right) < \varepsilon .
\end{equation}
We therefore have
\begin{eqnarray}
 \sum_{z\in\mathcal{S}}P(z,x)A(z,x;t)&=P(y,x)A(y,x;t)+
  \sum_{z\in\mathcal{S}\setminus\{y\}} P(z,x)A(z,x;t) \nonumber \\
 &<P(y,x)\varepsilon+\sum_{z\in\mathcal{S}\setminus\{y\}} P(z,x) \nonumber \\
 &=1-(1-\varepsilon)P(y,x),
\end{eqnarray}
and consequently,
\begin{equation}
 G(x,x;t)>(1-\varepsilon) P(y,x)>0.
\end{equation}
Since the right-hand side of \eref{eq:LB2} can be arbitrarily small for
sufficiently large $t$, we obtain the second part of Lemma \ref{lemma:1}.

\section{Proof of \eref{eq:MBD1} and \eref{eq:MBD2}}\label{appendix:B}

We use the following notations:
\begin{eqnarray}
 A(x)=\exp\left(-\frac{F_0(x)}{T_0}\right), \quad
 B=\sum_{x\in{\mathcal S}_1^{\rm min}} A(x), \\
 \Delta(x)=F_1(x)-F_1^{\rm min}.
\end{eqnarray}
If $x\in\mathcal{S}_1^{\rm min}$, the Boltzmann distribution can be
rewritten as
\begin{equation}
 q(x;t)=\frac{A(x)}
 {\displaystyle B +\sum_{y\in\mathcal{S}\setminus{\mathcal S}_1^{\rm min}}
 \exp\left(-\frac{\Delta(y)}{T_1(t)}\right)A(y)}.
\end{equation}
Since $\Delta(y)\geq 0$ by definition, the denominator decreases with
time. Thus, we obtain \eref{eq:MBD1}.

To prove \eref{eq:MBD2}, we consider the derivative of $q(x;t)$ with
respect to $T_1(t)$,
\begin{equation}\fl
 \frac{\partial q(x;t)}{\partial T_1(t)}=
 \frac{A(x)\left\{\displaystyle B \Delta(x)
  +\sum_{y\in{\mathcal S}\setminus \mathcal{S}_1^{\rm min}}(F_1(x)-F_1(y))
  \exp\left(-\frac{\Delta(y)}{T_1(t)}\right)A(y)\right\}}
 {\displaystyle T(t)^2 \exp\left(\frac{\Delta(x)}{T_1(t)}\right)\left[\displaystyle
 B+\sum_{y\in{\mathcal S}\setminus \mathcal{S}_1^{\rm min}}
 \exp\left(-\frac{\Delta(y)}{T_1(t)}\right)A(y)\right]^2}.
\end{equation}
Only $F_1(x)-F_1(y)$ in the numerator has the possibility of being
negative. However, the first term $B\Delta(x)$ in the curly brackets is
larger than the second one for sufficient large $t$ because
$\exp\left(-\Delta(y)/T_1(t)\right)$ tend to zero as
$T_1(t)\rightarrow\infty$. Thus there exists $t_1>0$ such that $\partial
q(x;t)/\partial T(t)>0$ for all $t>t_1$. Since $T_1(t)$ is a decreasing
function of $t$, we have \eref{eq:MBD2}.

\section{Proof of \eref{eq:q_GFMC}}\label{appendix:C}

The transition probability defined by \eref{eq:G}, \eref{eq:P_GFMC} and
\eref{eq:A_GFMC} is rewritten in terms of the weight \eref{eq:w-GFMC} as
\begin{equation}
 G_1(y,x;t)=\cases{
  1-\frac{N\Delta t\, \Gamma(t)}{w(x;t)} & ($x=y$) \\
  \frac{\Delta t\, \Gamma(t)}{w(x;t)} & 
 ($x\in\mathcal{S}_y$; single-spin flip) \\
  0 & (otherwise).
 }
\end{equation}
Thus, we have
\begin{eqnarray}
 \sum_{x\in\mathcal{S}} G_1(y,x;t)q(x;t) &=
 \left(1-\frac{N\Delta t\, \Gamma(t)}{w(y;t)}\right)\frac{w(y;t)}{A}
 +\sum_{x\in\mathcal{S}_y} \frac{\Delta t\, \Gamma(t)}{w(x;t)}
 \frac{w(x;t)}{A} \nonumber \\
 &=q(y;t)-\frac{N\Delta t\, \Gamma(t)}{A}
 +\frac{\Delta t\, \Gamma(t)}{A}\sum_{x\in\mathcal{S}_y}1,\label{eq:APP_C}
\end{eqnarray}
where $A$ denotes the normalization factor,
\begin{eqnarray}
 \sum_{x\in\mathcal{S}} w(x;t) 
 &= \Tr\left\{ 1-\Delta t\left(-\sum_{\langle ij \rangle}
 J_{ij}\sigma_i^z\sigma_j^z-E_T\right)
 +N\Delta t\, \Gamma(t) \right\} \nonumber \\
 &=2^N \left\{1+\Delta t\, E_T+N\Delta t\, \Gamma(t)
 \right\}.
\end{eqnarray}
Since the volume of $\mathcal{S}_y$ is $N$, \eref{eq:APP_C} indicates
that $q(x;t)$ is the stationary distribution of $G_1(y,x;t)$. The
right-hand side of \eref{eq:q_GFMC} is easily derived from the above
equation.


\end{document}